\documentclass[twocolumn, prl]{revtex4-1}
\usepackage{graphicx}
\usepackage{graphics}
\usepackage{subfigure}
\usepackage{slashbox}
\usepackage{amsmath}
\usepackage{brackets}
\usepackage{psfrag}
\usepackage{epsfig}
\usepackage{xcolor}
\usepackage{hyperref}
\usepackage{amssymb}
\usepackage{url}

\begin{document}

\ProvideTextCommandDefault{\textonehalf}{${}^1\!/\!{}_2\ $}

\title{Experimental search for an exotic spin-spin-velocity-dependent interaction using an optically polarized vapor and a rare-earth iron garnet}

\thanks{These authors have contributed equally to this work. }
\author{P.-H.~Chu*}
\email[Email address: ]{pchu@lanl.gov}
\author{Y.~J.~Kim*}
\email[Email address: ]{youngjin@lanl.gov}
\author{S.~Newman}
\author{I.~M.~Savukov}
\affiliation{MPA-Quantum, Los Alamos National Laboratory, Los Alamos, NM 87545, USA}
\author{C.~D.~Hughes}
\author{J.~C.~Long}
\affiliation{Department of Physics, Indiana University, Bloomington IN 47405
and IU Center for Exploration of Energy and Matter, Bloomington IN 47408, USA}
\date{\today}

\begin{abstract}
We report an experimental search for an exotic spin-spin-velocity-dependent interaction between polarized electrons of Rb atoms and polarized electrons of a solid-state mass, violating both the time-reversal and parity symmetries. This search targets a minute effective magnetic field induced by the interaction. A spin-exchange relaxation-free (SERF) magnetometer based on an optically polarized Rb vapor is the key element for both a source of polarized electrons and a high-sensitivity detector. A dysprosium iron garnet (DyIG) serves as the polarized mass, with an extremely small magnetization at the critical temperature around 240~K and a high spin density. To reduce the magnetization, one of major systematic effects, a home-built cooling system controls the mass temperature. To our knowledge, this is the first search for an exotic spin-dependent interaction using the compensated ferrimagnet DyIG as a polarized mass. The experiment set the most stringent limit on the electron-electron coupling strength in the centimeter interaction range, in particular $g_V^e g_V^e <10^{4}$ at $\lambda=2$~cm.
\end{abstract}

\maketitle
Experimental searches for exotic spin-dependent interactions between fermions are expanding, as part of the growing interest in the application of quantum sensing in high-energy physics~\cite{ahmed:2018quantum}. Early phenomenological work by Moody and Wilczek~\cite{Moody:1984} explored exotic spin-0 boson exchange, and was expanded by Dobrescu and Mocioiu~\cite{Dobrescu:2006au} to include potentials dependent on the relative velocity between interacting fermions through spin-1 boson exchange.  These bosons provide sensitive observables for testing theories beyond the Standard Model that can solve several outstanding mysteries in fundamental physics. For example, the strong charge-parity problem in quantum chromodynamics can be resolved by the axion~\cite{Peccei:1977}.  Cold dark matter can be composed of axions~\cite{Duffy:2009ig} or spin-1 dark photons ~\cite{Appelquist:2003,Dobrescu:2005,Ackerman:2009}. Theoretical solutions to the physics questions including the unification of gravity and the Standard Model, the hierarchy problem, and dark energy also predict the existence of new bosons~\cite{Safronova:2017xyt} which can mediate the exotic spin-dependent interactions.

Most experiments have focused on static spin-dependent interactions, described by the potentials $V_{2},~V_{3},~ V_{9+10}$ and $V_{11}$, where we have adopted the numbering convention in Ref.~\cite{Dobrescu:2006au}.  Techniques have included spectroscopy~\cite{Ficek_2017}, spin-polarized torsion pendulums, magnetometry, atomic parity non-conservation, and electric dipole moment searches~\cite{Safronova:2017xyt,Fadeev:2018rfl,Ritter:1990,Chui:1993,Ni:1994,Terrano:2015}. Some of these~\cite{Ritter:1990,Chui:1993,Ni:1994} used compensated ferrimagnets with low intrinsic magnetism as spin-polarized electron sources. A few searches for spin-velocity-dependent interactions between polarized and unpolarized fermions have been performed using cold neutron beams~\cite{Piegsa:2012,Haddock_2018}, magnetic stripes~\cite{Ding:2020mic}, and polarized $^{3}$He relaxation~\cite{Yan:2012wk,Yan:2015}. We also have conducted two such searches with a spin-exchange relaxation-free (SERF) magnetometer~\cite{Kim:2017yen,Kim:2019sry}.

Investigating spin-spin-velocity-dependent interactions (SSVDIs), on the other hand, is relatively challenging because the spin-polarized masses can generate a spurious magnetic signal, limiting the experimental sensitivity. Some experiments have been performed: Hunter {\it et al.} first applied polarized geoelectrons with a $^{199}$Hg-Cs co-magnetometer~\cite{Hunter:2014}; Ji {\it et al.} used a K-Rb SERF co-magnetometer with SmCo$_5$ spin sources~\cite{Ji:2018}; and at the atomic scale from an analysis of spin-exchange interactions~\cite{Kimball:2010}. However, while these experiments were sensitive at either large or extremely small distances, the medium interaction range is still lack of measurement. 

In this letter we explore a SSVDI at centimeter distances with a SERF magnetometer and a spin-polarized mass:
\begin{eqnarray}
V_{15}&=&-g_V^eg_V^e\frac{\hbar^3}{8\pi m_e^2 c^2 }\nonumber\\
&\times&\{[\hat{\sigma}_1\cdot(\vec{v}\times\hat{r})](\hat{\sigma}_2\cdot\hat{r})+(\hat{\sigma}_1\cdot\hat{r})[\hat{\sigma}_2\cdot(\vec{v}\times\hat{r})]\}\nonumber\\
&\times&\left(\frac{1}{\lambda^2 r}+\frac{3}{\lambda r^2}+\frac{3}{r^3}\right)e^{-r/\lambda},
\label{eq:v15}
\end{eqnarray}
where two interacting particles are electrons with spin unit vector $\hat{\sigma}_{1}$ inside the SERF vapor cell and $\hat{\sigma}_{2}$ in the mass, and the electron mass ($m_e$); their relative distance and relative velocity are $\vec{r}$ and $\vec{v}$; $\hbar$ is the reduced Planck constant; $c$ is the speed of light; and $\lambda$ is the interaction length. $g_V^e$ is the vector electron coupling~\cite{Leslie:2014mua, Fadeev:2018rfl} with the spin-1 dark photon~\cite{essig2013dark}. Note that $V_{15}$ violates both the time-reversal and parity symmetries. We also notice that some authors argued that the $V_{15}$ vanishes if both polarized fermions are identical~\cite{Fadeev:2018rfl}, but the corresponding constraints were still published~\cite{Hunter:2014,Ji:2018}. We investigated the $V_{15}$ based on our recently proposed experimental approach~\cite{Chu:2016}, where the SERF magnetometer serves as both a source of spin-polarized electrons and a high-sensitivity detector. The $V_{15}$ between $\hat{\sigma}_1$ with the gyromagnetic ratio $\gamma$ and $\hat{\sigma}_2$ generates an effective magnetic field $\vec{B}_{\text{eff}}$ that can interact with the SERF electron, similarly as the ordinary magnetic field~\cite{Karaulanov_2016}: $V_{15}=\Delta E=\gamma\hbar\hat{\sigma}_1\cdot\vec{B}_{\text{eff}}$, where $\Delta E$ is the energy shift of the SERF Rb spin-polarized electrons. The $B_{\text{eff}}$ is our signal to be measured with the SERF magnetometer. As the SERF magnetometer, we utilized a QuSpin cm-scale magnetometer based on an optically polarized $^{87}$Rb $3$~mm cubic vapor cell~\cite{QuSpin} (for more detail, see Refs.~\cite{Kim:2017yen,Kim:2019sry,Savukov_2017}). The cell was heated to about $\sim160~^\circ$C to elevate a Rb atomic density to $\sim10^{14}$~cm$^{-3}$. 

\begin{figure}[t!]
\centering
\includegraphics[width=0.45\textwidth]{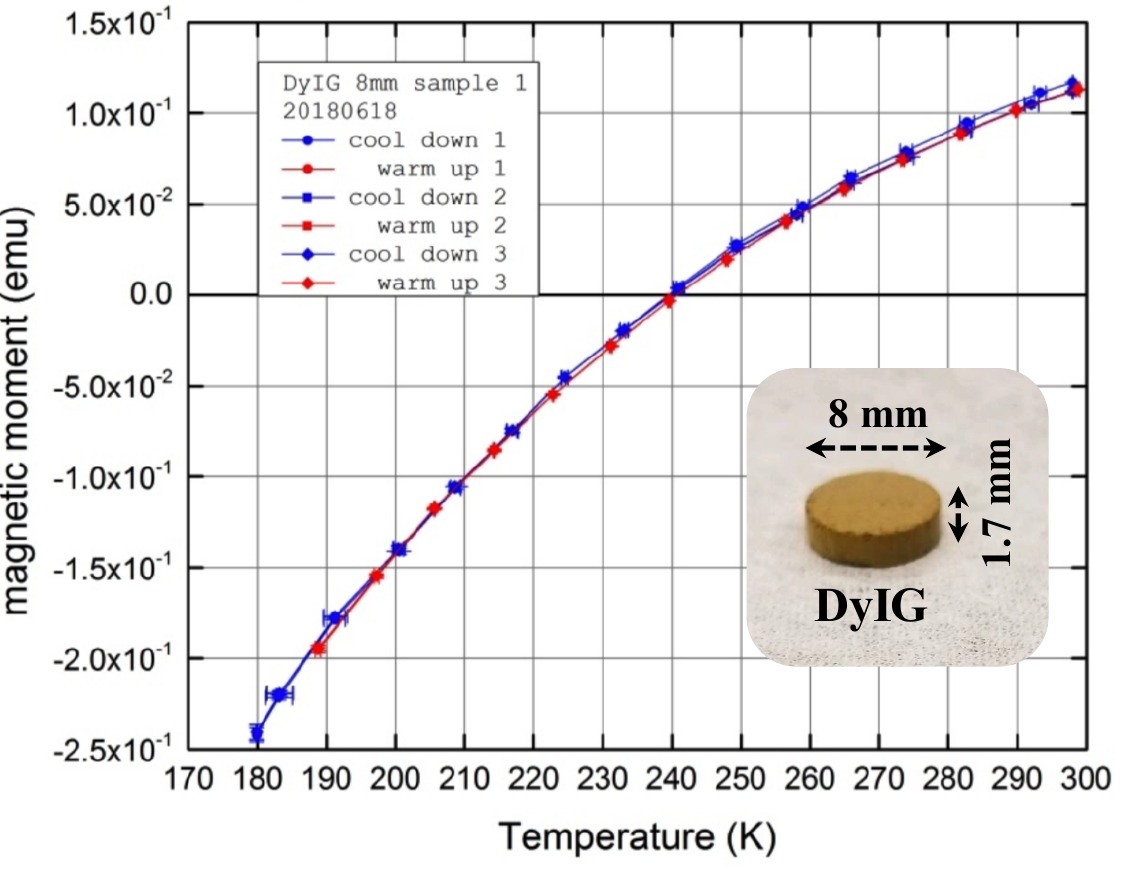}
\caption{The measured magnetism of the DyIG sample synthesized at Indiana university (the inset) as a function of the temperature. The measurement was performed by cooling  and warming  the sample in several cycles. 
}
\label{fig:DyIGmeasurement}
\end{figure}

A rare-earth iron garnet (dysprosium iron garnet, $\text{Dy}^{3+}_{3}\text{Fe}^{3+}_{2}\text{Fe}^{3+}_{3} \text{O}_{12}$, DyIG) was employed as the spin-polarized mass. DyIG is a ferrimagnet which exhibits orbital compensation of the magnetism due to the electron spins.  Three sublattices contribute to the net magnetization: Dy$^{3+}$ ions occupy dodecahedral sites in the garnet lattice, and Fe$^{3+}$ ions occupy octahedral and tetrahedral sites~\cite{garnet}. The magnetic moments of Dy$^{3+}$ are nominally aligned with the octahedral ion moments but anti-aligned with the tetrahedral ion moments. As in any ferrimagnetic material, DyIG exhibits net magnetization below the Curie temperature.  At the critical or compensation temperature ($T_c$) below the Curie temperature, the two opposing Dy$^{3+}$ and Fe$^{3+}$ moments equalize, resulting in a zero net magnetic moment and hence zero magnetization. Part of the Dy$^{3+}$ magnetism is orbital, so there is a spin excess at $T_c$, calculated to be 0.6 spins per molecule~\cite{Leslie:2014mua}. Based on the measured mass density of the sample used, the electron spin density of DyIG is $1.7\times10^{26}~\text{m}^{-3}$. To our knowledge, this is the first experiment to employ DyIG since it was proposed for spin-polarized masses~\cite{Leslie:2014mua}, due to its $T_c$ near room temperature and relatively high spin density. A polycrystalline DyIG sample was synthesized at Indiana University using a metal hydroxide precipitation technique~\cite{Leslie:2014mua, Geselbracht:1994}. The resulting sample had diameter 8 mm, thickness 1.7 mm and a mass of 0.32 g, as shown in the inset of Fig.~\ref{fig:DyIGmeasurement}.  The sample was then characterized in a SQUID magnetometer (Quantum Design MPMS-XL) ~\cite{MPMS}.  It was magnetized to saturation, after which the applied field was switched off and the remnant magnetization measured as a function temperature.  The results are shown in Fig.~\ref{fig:DyIGmeasurement}, which indicates a repeatable $T_c$ of 240 K.  Before installation in the experiment at Los Alamos National Lab (LANL), the sample was re-magnetized to saturation along its symmetry axis with a permanent magnet.

A dominant systematic effect in our experiment is the magnetic field generated by the spin-polarized DyIG sample. According to Fig.~\ref{fig:DyIGmeasurement}, this field can be strongly suppressed by keeping the sample at $T_c$, which we do with a cooling system as discussed below.  While previous experiments used ferrimagnets at room temperature to take advantage of the partial compensation~\cite{Ritter:1990,Ni:1994,Chui:1993}, to our knowledge this is the first experiment to attempt operation at $T_c$ for maximal cancellation.

\begin{figure*}[t!]
\centering
\includegraphics[width=0.9\textwidth]{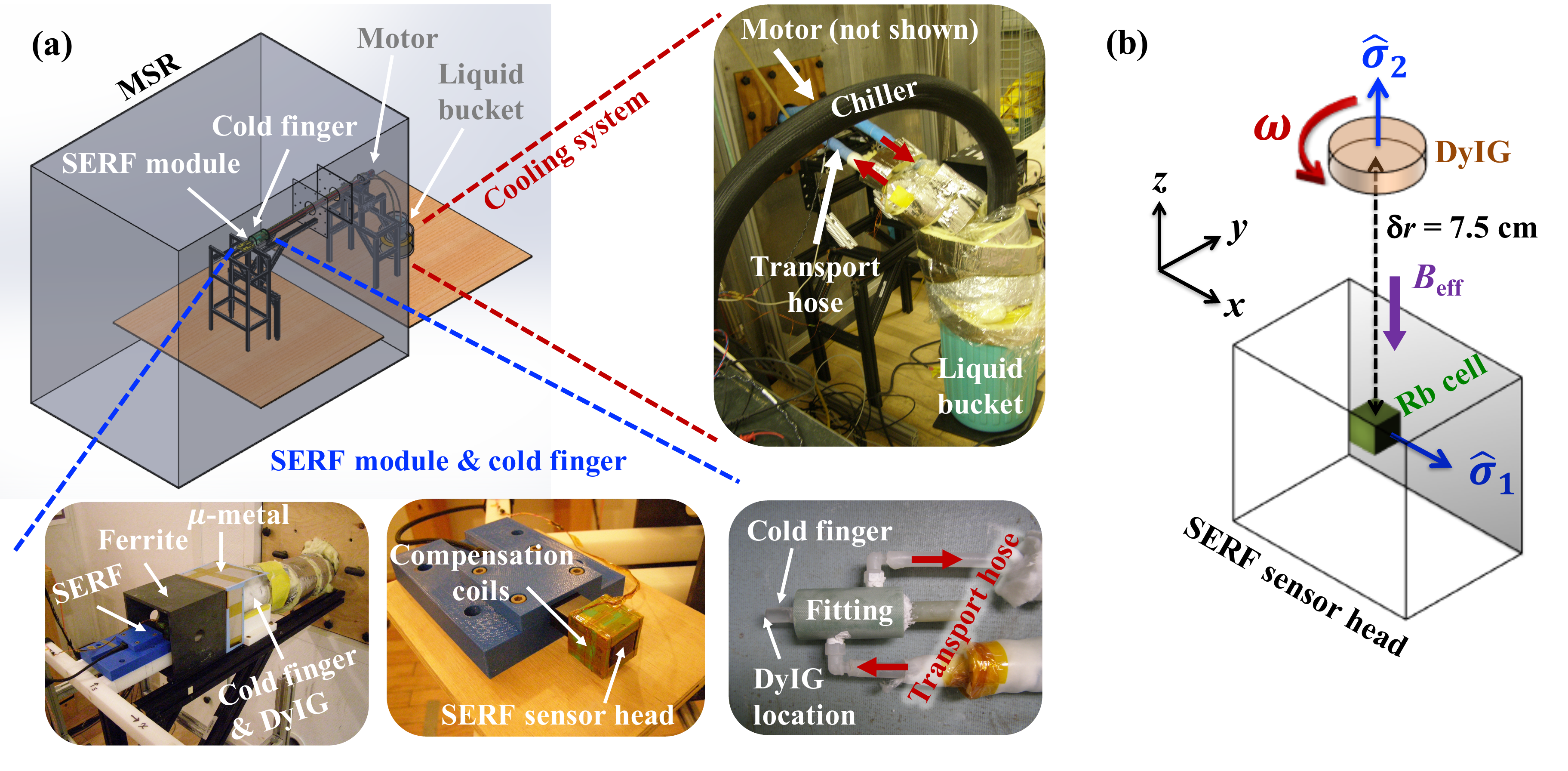}
\caption{(a) A schematic of the experimental setup comprised of a home-built cooling system and a SERF magnetometer module (scaled) and photos of main elements. The main elements of the cooling system, the chiller and the liquid bucket, are located outside the MSR. The cooled liquid in the bucket circulates through the transport hose connected to the plastic fitting containing the sapphire cold finger inside the MSR. The DyIG mass is attached at the end of the cold finger (not shown). The cold finger/mass assembly is wrapped by aerogel thermal insulation inside the plastic tube. The motor outside the MSR is connected to the assembly to rotate the mass. The SERF magnetometer with compensation coils is inside the MSR and its Rb vapor cell is concentrically aligned with the DyIG mass. The open ferrite box and one-layer $\mu$-metal box are located between the magnetometer and the mass. (b) A schematic of the configuration of the Rb vapor and the DyIG mass. The mass is rotated around the $z$-axis and the magnetometer sensitive to the field component along the $z$-axis measures $B_{\text{eff}}$.}
\label{fig:schematic}
\end{figure*}

Figure~\ref{fig:schematic}(a) shows a schematic of the experimental setup at LANL to probe $V_{15}$ and photos of main elements. Because the SERF magnetometer operates in low-field environments, it was located inside a magnetically shielded room (MSR), and also compensation coils were added at the magnetometer head to additionally cancel the fields from magnetic sources inside the MSR. The main body of the circulating cooling system--a chiller (PolyScience IP-100) that can achieve the temperature as low as 180~K in a liquid cryostat and an alcohol liquid bucket that contains a submersible pump for liquid circulation--was located outside the MSR in order to avoid their magnetic noise. For the design simplification, the cooling system components except the chiller were wrapped by flexible thermal insulation made of aerogel and fiberglass. In such an open environment, the chiller could cool the liquid in the bucket as low as 230~K. The liquid circulates through 2~m-long transport hose made of 1~cm-diameter PVC clear tubing that was connected to a plastic fitting placed inside the MSR through a hole with 6.35~cm radius on the MSR wall. The fitting contains a cold finger of a sapphire rod with 1~cm diameter and 5~cm length provided by Egorov Scientific. For the mass cooling, the DyIG mass was attached at the end of the sapphire rod with a high thermal conductivity (34.6~W/m/K) and a low magnetic susceptibility ($-2.1\times10^{-7}$; thus, no systematic magnetic signal is generated). The mass was concentrically aligned with the SERF Rb vapor cell. For the mass motion, a motor (Haydonkerk EC042B-2PM0-804-SP), located outside the MSR and enclosed by an one-layer $\mu$-metal box, was connected to the fitting through a G-10 rod. The cold finger/mass assembly was enclosed by a plastic cylindrical box filled with 5~mm-thick aerogel sheet.

A temperature sensor (Lake Shore DT-670-SD) was mounted on the transport hose near the liquid bucket to monitor the liquid temperature. It was observed that the temperature was around 235~K with the drift of 4~K for one day due to the ambient temperature variation. We observed that the mass temperature was a few degrees higher than the liquid temperature, thus  close to the $T_c$ of 240~K. Although the mass was cooled down to around the critical temperature, the residual field from the mass was measured to be $\sim1~\mu$T and the field drift caused by the temperature drift deteriorated the performance of the SERF magnetometer. To this end, the cold finger/mass assembly was surrounded by an open thin one-layer $\mu$-metal box and an open ferrite box additionally enclosed the SERF magnetometer, as shown in Fig.~\ref{fig:schematic}(a). This configuration resulted in the increase of the distance between the nearest surfaces of the Rb vapor cell and the DyIG mass, $\delta r$, up to 7.5~cm.

Figure~\ref{fig:schematic}(b) illustrates a schematic of the configuration of the SERF Rb vapor with spins oriented along the $x$-axis and the DyIG mass with spins oriented along the $z$-axis. In order to generate the relative velocity term in $V_{15}$, the mass was rotated by the motor with a constant angular velocity $\omega$, leading to the $B_{\text{eff}}$ along the $z$-axis that can be precisely measured by the SERF magnetometer, sensitive to the $z$ field component with the intrinsic field sensitivity of 15~fT/$\sqrt{\text{Hz}}$ at low frequency below 100~Hz.

\begin{figure}[t!]
\centering
\includegraphics[width=0.5\textwidth]{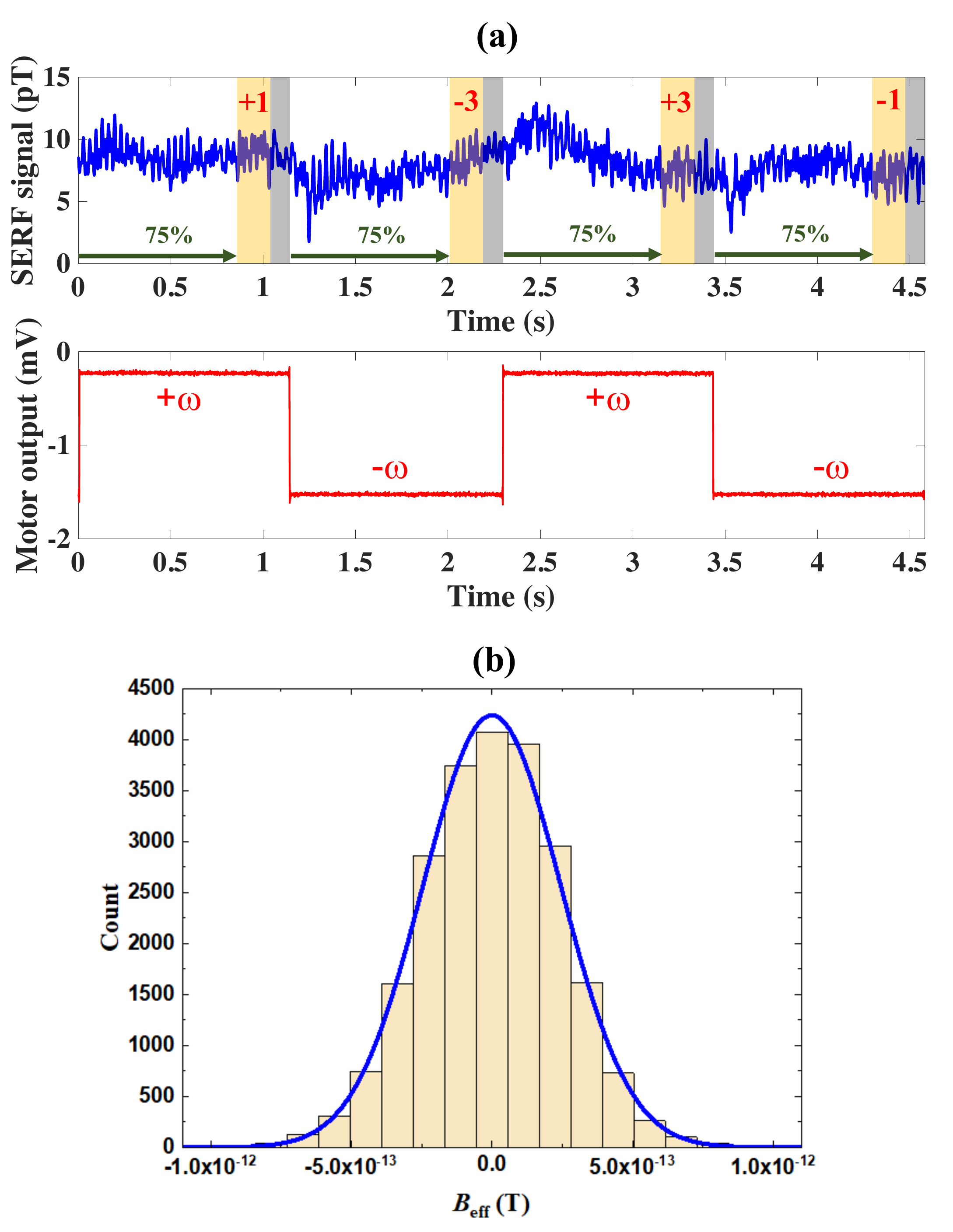}
\caption{(a) SERF magnetometer signals (top) and motor trigger signals (bottom) in the time domain. (b) Histogram of the $B_{\text{eff}}$ extracted with the drift-correction algorithm from the data collected for 29.4~h; the blue curve represents a fit to a Gaussian distribution. }
\label{fig:signal}
\end{figure}

The suppression of the systematic effects (SERF dc offset on the order of pT; mass residual field on the order of $\mu$T) was achieved by continuously alternating the mass rotation between clockwise and counterclockwise to subtract the magnetometer signals because the sign of $B_{\text{eff}}$ is reversed for the opposite rotations, unlike the systematic effects. The motor outputed a trigger signal indicating the rotation direction, enabling to distinguish the direction in the magnetometer signals. Figure~\ref{fig:signal}(a) shows standard magnetometer signals in the time domain together with motor trigger signals, both of which simultaneously recorded, that represent two full cycles of the mass rotation reversal. In one cycle, the mass was rotated with $\omega=0.242$~rad/s for 1~s and then with $\omega=-0.242$~rad/s for 1~s. Due to the acceleration/deceleration times and the delay time after each mass rotation in the motor, one cycle elapsed 2.3~s. 

To obtain the magnitude of $B_{\text{eff}}$ from the magnetometer signals, only data points within the yellow shaded regions (the last $25\%$ of data without the regions of the motor deceleration and delay, marked as the gray shaded regions) in each half cycle were used [see Fig.~\ref{fig:signal}(a)] in order to diminish the effects originating from motor acceleration such as mass vibration and also to ensure stable mass rotation with the constant angular velocity. The magnetometer signals between the opposite mass rotations were effectively subtracted using the drift-correction algorithm (for more detail see Refs.~\cite{Kim:2017yen,Kim:2019sry}). The algorithm removes drifts mainly due to the mass field drift (2~nT per one day) up to second-order time-dependent terms in the signals  by applying a [$+1~-3~+3~-1$] weighting to the mean value of the data within the yellow shaded regions from each half cycle within two cycles [see Fig.~\ref{fig:signal}(a)].

We collected data for 29.4 hours. Figure~\ref{fig:signal}(b) shows a histogram of the magnitude of $B_{\text{eff}}$ obtained with the drift-correction algorithm from the  magnetometer signals recorded for 29.4 h, which was fit with a Gaussian distribution, giving $B_{\text{eff}} = (-0.49\pm1.61)\times 10^{-15}$~T.  The dominant systematic effects have been mitigated below the statistical sensitivity of $1.61\times 10^{-15}$~T corresponding to the experimental sensitivity $\Delta E$ of $2.92\times 10^{-19}$~eV.

\begin{figure}[t!]
\centering
\includegraphics[width=0.46\textwidth]{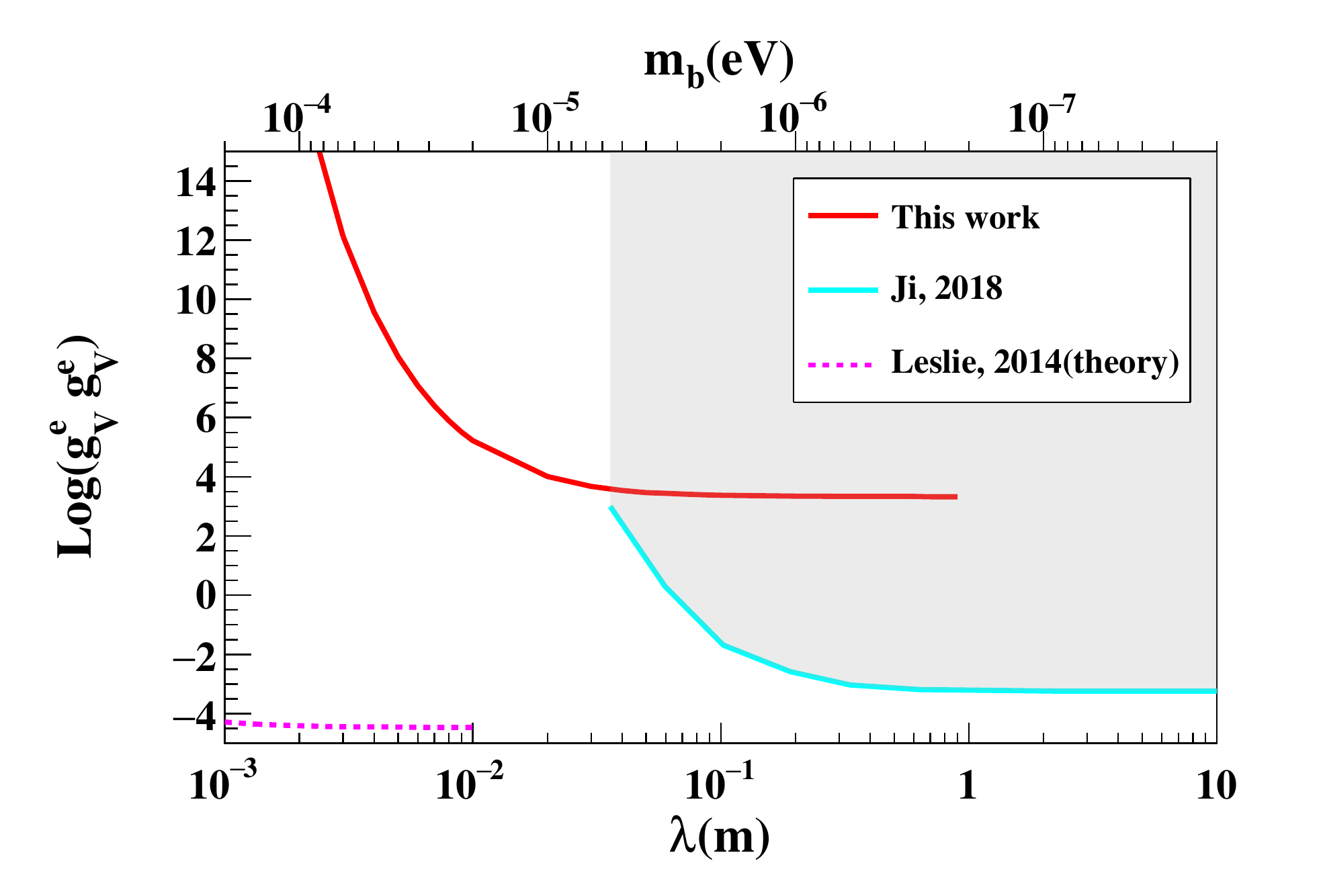}
\caption{Experimental limits on the electron-electron coupling $g_V^e g_V^e$ of $V_{15}$ as a function of the interaction range and the dark photon mass $m_b$ of our experiment (red solid curve) and the K-Rb SERF experiment~\cite{Ji:2018} (blue solid curve). The limit derived from the $^{199}$Hg-Cs experiment~\cite{Hunter:2014} is not shown. The magenta dashed curve shows an estimated limit of another proposal using DyIG~\cite{Leslie:2014mua}.}
\label{fig:copuling_1}
\end{figure}

The limit to $g_V^e g_V^e$ of the interaction $V_{15}$ was derived using the Monte Carlo method to average the interaction potential in Eq.~\ref{eq:v15} at different interaction ranges (for more detail, see Refs.~\cite{Chu:2016,Kim:2017yen,Kim:2019sry}), plotted in Fig.~\ref{fig:copuling_1} (red solid curve). The other experimental constraints on $g_V^e g_V^e$ have been derived from the experiments based on the $^{199}$Hg-Cs co-magnetometer with polarized geoelectrons~\cite{Hunter:2013, Hunter:2014} (not shown) and the K-Rb SERF magnetometer with SmCo$_5$ spins~\cite{Ji:2018} (blue solid curve) for the long interaction range $>1$~m. Our experiment sets a new limit on $V_{15}$ in the interaction range from $10^{-3}$ to $10^{-1}$~m. 

In conclusion, we probed the exotic parity- and time-reversal-odd SSVDI $V_{15}$ between SERF spin-polarized electrons and DyIG spin-polarized electrons, and set the most stringent constraint on the electron-electron coupling strength at the centimeter interaction range. The result indicates that this experiment is able to explore the remaining SSVDIs by proper mass movements. The interaction range in this experiment was limited by the field drift of the spin-polarized mass. For a shorter interaction range, a reduction of the mass field drift should be achieved by developing a vacuum cooling system with a thermal feedback loop. 

This work was supported by the Los Alamos National Laboratory LDRD office through Grant No. 20180129ER, the National Science Foundation grant PHY-1707986, and the Indiana University Center for Spacetime Symmetries (IUCSS). We are grateful to Alex Brown, who passed away in 2018, for his work on the synthesis of the DyIG sample.

\bibliography{main}
\end{document}